\documentclass[preprint,draft,aps]{revtex4}
\usepackage{graphicx,amssymb}

\begin{document}
\title{Long wavelength behavior of two dimensional photonic
crystals}
\author{S. T. Chui$^1$ and Zhifang Lin$^2$}
\affiliation{$^1$Bartol Research Institute, University of Delaware, Newark, DE\\
19716, $^2$Physics Dept., Fudan University, Shanghai, China.}

\begin{abstract}
We solve {\bf analytically} the
multiple scattering (KKR) equations for the 
two dimensional photonic crystals in the long wavelength limit.
Different approximations of the electric
and magnetic susceptibilities are presented from a unified
pseudopotential point of view.
The nature of the so called plasmon-polariton bands are clarified.
Its frequency as a function of the wire radius is discussed.

\end{abstract}
\maketitle
There is much interest recently in two dimensional (2D)
photonic crystals (PC) consisting of arrays of metallic 
or dielectric cylinders (wires)
in an insulating matrix or arrays of insulating cylinders in a metallic matrix.
These include recent interest in left-handed materials\cite{Pendry}
and in plasmonics.\cite{plasmonics}
A key issue is the effective susceptibilities
$<\epsilon>$ and $<\mu>$ of the system.
To design systems at different frequencies such as in the 
infrared range it is useful to know their values for different
system parameters.
The photonic bands in an array of
cylinders can be understood \textbf{entirely} in terms of the
scattering phase shift of the cylinders.
In the pseudopotential idea in electronic structure calculation 
a real potential is replaced by an effective one so that 
the same scattering of the electrons is produced.
Similarly effective susceptibilities can be introduced so that the
correct scattering effect for electromagnetic waves is produced.
We examine this idea to derive effective susceptibilities
for the cylinder. For example, from the scattering
phase shift for a mode with the electric field along the cylinder axis,
an effective dielectric constant for the cylinder is found to be
\begin{equation}
\epsilon_E'= -2J'\epsilon/(Jk_iR)[1 +0.5(k_oR)^2\ln(k_oR)]
/[1-\mu_o k_iR J' \ln(k_oR)/(J\mu_i) ];\ 
\end{equation}
$\epsilon=\epsilon_i/\epsilon_o,$ $J=J_0(k_iR)$.
The subscipts i, o refers to quantities inside and outside the cylinder, 
respectively. For metallic cylinders, when the skin depth is much less
than the radius of the cylinder, the second term in the 
denominator is larger than the first term, we obtain an effective
dielectric constant of a metallic form given by
$$
\epsilon_E' = 1-\omega_p^{\prime\ 2}/\omega^2
$$
where the effective "plasma frequency" is given by
\begin{equation}
\omega_p^{\prime\ 2}=-2c^2/[R^2 \ln(\omega R/c)];
\end{equation}
$R$ is the radius
of the cylinder.  While the original analysis\cite{Pendry_wire}
for the effective dielectric constant is carried out for a wire
radius less than the skin depth, the experiments\cite{Smithapl}
for the left-handed
materials are carried out for wires the width of which is larger than the 
skin depth. The above formula provides for an extension of the original
analysis.  There is a log correction term that has not been 
noticed before.

This clarifies the issue of damping. For frequencies from 1 to 10 GHz,
the imaginary part of the dielectric constant of most metals is about a
thousand times larger than the real part.
When the skin depth is much less
than the wire radius, the loss
in the metal is only restricted near the surfaces of the wires and
the effective damping is reduced.
Indeed, the above effective dielectric constant 
depends only on the wavelength and the wire radius, with no damping!
 
Eq (1) encompasses other recent results.
For dielectric rods. If 
the second term of the denominator is smaller than the first term, 
we recover recent results
by Wu et al. and Hu et al.\cite{wu,hu} that 
$$
\epsilon_E'= -2J'\epsilon/(Jk_iR) 
$$
This pseudopotential idea is also implicit in recent results using cylinders
with a high dielectric constant ferroelectric.\cite{peng}
Eq. (1) extends these results to more general regions of the parameter space. 

In this paper we further calculate the photonic band structure 
of an array of cylinders of radius R 
in the long wavelength limit when the separation between the wires $a$ 
is less than the free space wavelength $\lambda=2\pi/k_0$.  
We solve analytically the multiple scattering (KKR) equations
in the long wavelength limit.
We find that both the s and the p wave scattering phase shifts are
of the same order of magnitude, $(k_0R)^2$, and need to be considered.
These produced {\bf two} photonic branches, an "acoustic" mode 
with a frequency proportional to the wave vector with an effective 
dielectric constant $<\epsilon>$ (eq. (1) and (8))
and a magnetic susceptibility $<\mu>$ (eq. (5) and (8))
and an "optic" mode with a gap.
For {\bf negative} susceptibilities and narrow cylinders, 
the "optic" mode corresponds to a flat band at frequencies
close to the surface plasmon resonances, as has been previously discovered numerically.
For the acoustic mode, we found that $<\epsilon>$ can be expressed as the
arithmetic mean of that of the medium and an effective dielctric constant
of the cylinder $\epsilon_c'$.  We now describe our result in detail.

{\bf Pseudopotential:}
We first describe our "pseudopotential" idea for the effective dielectric
constant of the cylinder.
As far as the EM field outside the cylinder is concerned, all that matters
is the scattering phase shift for angular momentum component n given, for the E
(TM) mode, by
\begin{equation} \label{bn}
\tan\eta_n^E =\frac{ J'_n(k_oR)k_iJ_n(k_iR) -{\it J'_n(k_iR)} J_n(k_oR)k_o\epsilon }{
  k_iJ_n(k_iR) {\it N'_n(k_oR)-k_o\epsilon N_n(k_oR){\it J'_n(k_iR)}
}}
\end{equation}
$k_j=k_0(\mu_j\epsilon_j)^{1/2}$ for j=o,i,
$\epsilon=\epsilon_i/\epsilon_o$.
Similarly, for the H (TE) mode
$k_{i}=(\epsilon_i\mu_i)^{1/2}k_0$.
\begin{equation}
\tan\eta_n^H = \frac{ J'_n(k_oR)k_o\epsilon J_n(k_iR) - J'_n(k_iR) J_n(k_oR)k_i}{
  k_o\epsilon J_n(k_iR)  N'_n(k_oR)-k_i N_n(k_oR) J'_n(k_iR)
}.
\end{equation}
We first focus on the s wave with n=0.  For $k_oR<<1,$
$J_0=1,$ $J_0'=-k_oR/2,$ $N_0=2 ln(k_oR)/\pi,$ $N_0'=2/(\pi k_oR).$
$$\tan\eta_0^E\approx -0.25(k_oR)^2\pi
[1+2J'\epsilon/(Jk_iR) ]/[1-\mu_o k_iR J' \ln(k_oR)/(J\mu_i) ],$$
$$\tan\eta_0^H =-0.25(k_oR)^2\pi [ 1+ 2J' k_i/(J\epsilon k_o^2R)]/[
1- k_iR \ln(k_oR) J'/(J\epsilon)].$$

For $k_iR$ also small
$$
\tan\eta^E_0 \approx -\pi (k_oR)^2[ 1  - \epsilon ] / 4 ; \   
\tan\eta^H_0 \approx -\pi (k_oR)^2[ 1  - \mu_i/\mu_o] / 4  
$$
As is expected, when $\epsilon=1,$ there is no scattering and
$\tan\eta_0^E=0.$

When $k_iR$ is not small, one can {\bf define} effective 
susceptibilities so that the same phase shift is produced:
\begin{equation}
\tan\eta^E_0 \approx -\pi (k_oR)^2[ 1  - \epsilon_E' ] / 4; \   
\tan\eta^H_0 \approx -\pi (k_oR)^2[ 1  - \mu_H'/\mu_o] / 4  
\end{equation}
This is the "pseudopotential" idea that we mentioned.
From eq. (5) we obtain eq. (1) for the effective dielectric constant and 
also an effective magnetic susceptibility for the H mode:
\begin{equation}
\mu_H'=-[  k_iR \ln(k_oR) J'/(J\epsilon)+ 2J' k_i/(J\epsilon k_o^2R)]/[
1- k_iR \ln(k_oR) J'/(J\epsilon)].
\end{equation}

The second term in the denominator is of the order of $(\mu_i
/\epsilon_i)^{0.5} k_oR$ and is usually smaller than the first term.
We obtain
$$
\mu_H'\approx  - 2J' k_i/(J\epsilon k_o^2R).
$$


We next investigate the phase shifts for the higher order partial waves.
In the limit $k_oR<<1$, 
$$\tan\eta_n^E =\pi (k_oR/2)^{2n}/((n-1)!n!)
[ \mu_o-\mu_i nJ_n/(J_n'k_iR)]/[\mu_o+ \mu_i nJ_n /(J_n'k_iR)],$$
$$\tan\eta_n^H = \pi (k_oR/2)^{2n}/[(n-1)!n!]
[ 1-n\epsilon J_n/(J_n'k_iR)]/[ 1+ n\epsilon J_n  /(J_n'k_iR)].$$
Here $J_n=J_n(k_iR).$
When $k_iR$ is also small
$$\tan\eta_n^E =[(k_oR/2)^{2n}/n](\mu -1 )/(\mu+1);\ 
\tan\eta_n^H =[(k_oR/2)^{2n}/n](\epsilon -1 )/(\epsilon+1).
$$
There is recently much interest in "plasmonics"
when the frequency is close to the interface plasmon frequency
so that $\epsilon=-1.$.
At this frequency $\eta_n^H=\pi/2$.
A Mie scattering resonance is exhibited for the TE 
modes {\bf for all $n\neq 0$}.
For $n=1$
the requirement that the same scattering phase shift is
obtained even when $k_iR$ is not small
provides for the equations determining the effective 
susceptibilities:
\begin{equation}
\tan\eta_1^E =-\pi (k_oR/2)^{2}(\mu_o-\mu_E')/(\mu_E'+\mu_o).
\end{equation}
\begin{equation}
\tan\eta_1^H = -\pi (k_oR/2)^{2}(\epsilon_o-\epsilon_H' )/
(\epsilon_o+ \epsilon_H' ).
\end{equation}
From these we obtain the effective susceptibilities
\begin{equation}
\mu_E'=\mu_i J_1/(J_1'k_iR);\ 
\epsilon_H'=\epsilon_i J_1/(J_1'k_iR).
\end{equation}
Similar equations have also been obtained by Hu et al. 
and Wu et al.\cite{hu,wu} from a coherent potential approximation.
The results here provides a different interpretation of their results.
With the current view, "plasmonics" phenomena can also be manifested
for non-metallic rods if $\epsilon_H'+\epsilon_o=1$ and
the same scattering phase shift is produced.
We next turn our attention to the photonic bands.

{\bf Photonic band structure:}
First we briefly recapitulate the multiple scattering (KKR) technique.
The basic idea is that the scattered wave from the photonic
crystal is self-consistently
sustained. More precisely, consider a cylinder at the origin. The 
scattered wave from all the {\bf other} cylinders 
sum to produce a net incoming wave at the origin
which is scattered by the cylinder at the origin and in turn produce a
scattered wave from the origin. This scattered wave from the origin is related
to the scattered wave from the other sites by a phase factor determined
by the wave vector. 

More precisely, we denote the amplitude of the partial scattered wave
with angular momentum n
by $a_n.$ The sum of the scattered waves from all the {\bf other} sites
becomes an incoming wave at the orgin with the amplitude 
$p_{n}=\sum_n a_{n'} S(n-n')$ where 
\begin{equation}
S(m)=
\sum_{R\neq 0}exp(i{\bf k}\cdot \mathbf R) H_m(k_oR) \exp(im\phi_{R}).
\end{equation}
Note that
$S$ does {\bf not} include the wave from the origin; thus the sum in eq. (10)
does not include the term at R=0. The outgoing scattered wave at the origin
is related to the incoming wave by the t matrix: $a_n=t_np_n.$ Substitute
in the definition of $p_n$, we arrive at the equation $det[S(n-n')-\delta(n-n')
/t_n]=0.$ Since the t matrix is related to the phase shift by
$t_n=\tan\eta_n/(\tan\eta_n+i),$ we obtain the KKR equation
\begin{equation}
det[A(n-n')-\delta(n-n') \cot\eta_n]=0.
\end{equation}
Here the structure factor $A(m)=[S(m)-1]/i.$ 
In this paper, we have assumed a time dependence of $\exp(i\omega t).$
Outgoing spherical waves correspond to a $H_n=J_n+iN_n.$

In the long wavelength limit, one can approximate the sum
for $S$ by an integral, which can then be analytically evaluated.\cite{detail}
The structure factor becomes
$$A(n)\approx 4 [i\exp(i\phi_k)k/k_o]^n /[(k_oa)^2(k^2/k_o^2-1)] $$
As is discussed above,
if the wavelength outside the cylinder is long and $k_oR<<1,$
$\tan\eta_n\propto (k_oR)^{2n}$ for $n\neq 0,$
$\tan\eta_0\propto (k_oR)^{2}.$ The s and the p wave shifts
are of the same order of magnitude, $(k_oR)^2$, and need to be considered.
When only the s and the p wave components are kept,
the KKR equation now becomes
\begin{equation}
{\bf H E}=0
\end{equation}
where
$${\bf H}=\left [
\begin{array}{cccc}
A(0)+\cot\eta_1 & A(1) & A(2) \\
A(1)^*& A(0)+\cot\eta_0& A(1) \\
A(2)^* & A(1)^* & A(0)+\cot\eta_1\\
\end{array}
\right ],
$$
There are two classes of solutions,
with either $E_1=E_{-1}^*=|E_1|\exp(i\phi_k)$ or
$E_1=-E_{-1}^*=i|E_1|\exp(i\phi_k).$
We get two possible eigenvalue equations. The first one is given by
\begin{equation}
[A(0)+\cot\eta_1+|A(2)|][A(0)+\cot\eta_0]-2|A(1)|^2=0.
\end{equation}
For the second case, we get
\begin{equation}
A(0)+\cot\eta_1+|A(2)|=0.
\end{equation}
As we show below, the first  mode corresponds to an "acoustic" branch
with a frequency proportional to the wave vector with effective susceptibilities
for the system; the second mode corresponds to a band with a gap.
For negative susceptibilities, this corresponds to a flat band at frequencies
close to the surface plasmon resonances, as has been previously discovered 
numerically \cite{Maradudin}.

We discuss the acoustic branch first.
%
%
Substituting in the expressions for the phase shifts (eq. (5) and (7))
and the structure factor into eq. (11) and after some algebra, we obtain
$$k^2=k_0^2<\epsilon><\mu>$$
where
\begin{equation}
<\epsilon> =\epsilon_o(1-f)+f\epsilon_i,
\end{equation}
\begin{equation}
<\mu>=\mu_o [\mu_i'(1+f)+\mu_o(1-f)]/[\mu_i'(1-f)+\mu_o(1+f)].
\end{equation}

In the static (zero
wavevector and frequency) 
limit\cite{Born}
for the case with the E field  along the axis 
the tangential component of the electric field in the
cylinder (i) and the
medium outside (o) is the same: $E_o=E_i=E.$ The average displacement field is
given by $<D>=c_oD_o+(1-c_o)D_i.$ Since $D_o=\epsilon_oE_o,$
$D_i=\epsilon_iE_i,$ we obtain
$<D>=(c_o\epsilon_o+(1-c_o)\epsilon_i)E=<\epsilon>E.$ Hence
the average dielectric constant is just the arithmetic mean of the 
dielectric constants of the components: 
$<\epsilon>=(c_o\epsilon_o+(1-c_o)\epsilon_i).$
This is the same as eq. (15).

In multilayer systems, a similar result is obtained.\cite{jpcm} In that case
the effective $\mu$ is the harmonic mean of the components while the
effective dielectric constant is still the arithmetaic mean of that
of its components.

We next discuss the "optic" mode.
Substituing in the expressions for the phase shifts and the A's,
the equation for the second optic mode becomes
$$ 2\ln \left( k_oa/2\sqrt{\pi} \right)   
 (k_oa)^2/\pi =4-4f^{-1}(\mu_o+\mu_E')/(\mu_E'-\mu_o)+O(k^2) $$
for the E mode and
$$ 2\ln \left( k_oa/2\sqrt{\pi} \right)   
(k_oa)^2/\pi =4-4f^{-1}(\epsilon_o+\epsilon_H')/(\epsilon_H'-\epsilon_o)+O(k^2)$$
for the H mode. When $k$ approaches zero, $k_o$ is not zero!
Let us illustrate the physics by looking at the H mode. The limit of small 
f is particularly interesting. In that limit, the frequency is determined
by the condition that $\epsilon_o+\epsilon_H'=0$ where $\epsilon_H'$
is given in eq. (8). 

For metallic cylinders with their radii less than the skin depths,
$k_iR<<1$,  $\epsilon_H'=\epsilon_i=
1-\omega_p^2/\omega^2$ where $\omega_p$ is the plasma frequency. 
$\epsilon_o+\epsilon_i=0$ when $\omega$ is equal to the inteface plasmon
resonance, $\omega_{sp}=\omega_p/(1+\epsilon_o)^{1/2},$ 
For small k, from the above equation, we see that $\omega(k)=
\omega(k=0)+O(fk).$ If $f$ is small, the dispersion is weak and the bands
are flat.  This flat band has been observed
numerically previously.\cite{Maradudin} The present calcualtion provides for a more direct
analytic demonstration of this result.
If $k_iR$ is not small, eq. (8) suggests that even with insulating
cylinders, flat "plasmonic" photonic bands can still be obtained if the condition
$$
\epsilon_i J_1/(J_1'k_iR)=-\epsilon_o.
$$
is satisfied.

Let us next look at the E mode, the condition becomes $\mu_E'/\mu_o=-1.$
We call this the "magnetic surface plasmon" mode.
Even though there is a lot of interest in plasmonics that focus on
the condition $\epsilon_H'/\epsilon_o=-1$, the corresponding condition
on $\mu$ have not been much discussed.

There is another way to think of this type of solutions.
As can be seen from eq. (5) and (6),
when the susceptibilities of the metal are negative,
$\epsilon_o+\epsilon_m$ can become zero and $\tan\eta=\infty.$ 
The scattering can go though resonances due to the interface plasmon. This
can lead to flat photonic bands, as has been observed in previous 
numerical calculations. In general, the more rapidly varying the phase shift,
the flatter the band.
 
Pokrovsky and Efros\cite{Efros} have recently investigated 
the propagation of electromagnetic (EM) waves in a periodic array 
of metallic cylinders (wires) in the limit $\kappa R>>1.$ Our conclusion 
differs from theirs.
In their paper, an expression similar to our $S(0)$ also appears. However,
in their expression, the sum is over {\bf all} R whereas the R=0
term is absent in ours.

This research is partly supported by the DOE.


\begin{references}
\bibitem{Pendry} J. B. Pendry, Phys. Rev Lett. {\bf 85}, 3966 (2000);
R. A. Shelby, D. R. Smith, and S. Schultz,
              Science {\bf 292}, 79 (2001).
\bibitem{Pendry_wire}
J. B. Pendry, A. J. Holden, D. J. Robbins and W. J. Stewart,
Jour. Phys. Conds. Matt., 10, 4785 (1998).
\bibitem{Smithapl}
R. A. Shelby, D. R. Smith, S. C. Nemat-Nasser and S. Schultz,
Appl. Phys. Lett., 78, 489 (2001). The dimensions of the wires are 0.25 mm x
\bibitem{Born}
M. Born and E. Wolf, Principles of Optics, 7th Ed., Cambridge 
University Press, (1999), p. 837.
0.03 mm x 1 cm.
\bibitem{plasmonics}
A.V. Zayats, I.I. Smolyaninov, A.A. Maradudin, "Nano-optics of surface
plasmon polaritons," Phys. Rep., vol. 408, pp. 131-314 (2005).
\bibitem{efm}
Xinhua Hu, C. T. Chan, Jian Zi, Ming Li and Kai-Ming Ho, Phys. Rev. Lett.
96, 223901, (2006); Ying Wu, Jensen Li, Zhao-Qing Zhang and C. T. Chan,
Phys. Rev. B 74, 085111 (2006).
\bibitem{detail2}
We expect $|k_iR|>2\pi R/\delta$ where the skin depth $\delta=c(2\pi \mu\omega
\sigma)^{-1/2}.$
In terms of the wavelength, we get $\delta=[c\lambda/(\mu\sigma)]^{1/2}/(2\pi).$
Take the conductivity of Cu, $\sigma=5.88\times 10^5\times 9
\times 10^{11} 1/sec,$ $\delta=0.38\times 10^{-4}(\lambda (cm)/\mu)^{1/2}.$
The smallest R possible with lithography is 0.1$\mu m.$ 
For $\lambda=1\mu m=10^{-4}cm,$ $|k_iR|>100 \mu_i^{1/2}.$
\bibitem{jpcm}
S. T. Chui, Z. F. Lin and C. T. Chan, Jour. Phys. Conds. Matt. 
18, L89 (2006).
\bibitem{Maradudin} 
A.V. Zayats, I.I. Smolyaninov, A.A. Maradudin, "Nano-optics of surface
plasmon polaritons," Phys. Rep., vol. 408, pp. 131-314 (2005).
\bibitem{Efros}
A. L. Prokrovsky and A. L. Efros, Phys. Rev. Lett., 89, 093901 (2002).
\bibitem{wu} Y. Wu, J. S. Li, Z. Q. Zhang, and C. T. Chan,
Phys. Rev. B {\bf74}, 085111 (2006). $\epsilon'$ here corresponds to
${\tilde \epsilon}_s$ in this paper.
\bibitem{hu} X. H. Hu, C. T. Chan, J. Zi, M. Li, and K. M. Ho, 
Phys. Rev. Lett. {\bf96}, 223901 (2006). See eq. (4) and (5) in this
paper.
\bibitem{peng} L. Peng, L. X. Ran, H. S. Chen, H. F. Zhang, J. A. Kong,
and T. M. Grzegorczyk, Phys. Rev. Lett. {\bf98} 157403 (2007).
\bibitem{detail} We change the variable R to x=kR and get
$$S(0)=\sum_{x\neq 0}(\Delta x)^2\exp(ik\cdot x/k_o) H_0(x)/(k_oa)^2 $$
In the long wavelength limit, $\Delta x$ becomes small.
We approximate this sum by an integral and get
$$S(0)\approx \int_{x_i}^{\infty}d^2x \exp(ik\cdot x/k_o) H_0(x)/(k_oa)^2. $$
We pick the lower limit so that the empty area remains the same.
($\pi x_i^2=(k_oa)^2$)
Since $\exp[ia\cos(\theta)]=\sum_m i^m J_m(a)\exp(im\theta),$ 
only the m=0 term remains in the integral. The radial integral can be 
easily done.
[We assume that $k_0$ has a small imaginary part so that the upper limit 
contribution can be set to zero.]
We get,  
$$S(0)\approx -2\pi 
x[d/dx(J_0(xk/k_o))H_0(x)-J_0(xk/k_o)d/dx(H_0(x))]/[(k_oa)^2(k^2/k_o^2-1)]|_{x_i} 
$$
Using the small argument expansions for the Bessel functions, we obtain
In the limit $x<<1$, $J_0(x)\approx [1-(x/2)^2]$, 
$J'(x)\approx -x/2+O(x^3)$,
$N_0(x)=2\ln(x/2)[1-x^2/4]/\pi.$
$N_0'(x)=2[(1-x^2/4)/x-x\ln(x/2)/2]/\pi.$
$$S(0)\approx  1+4i[1-x_i^2 (1-(k/k_o)^2/2)\ln x_i]/[(k_oa)^2 (k^2/k_o^2-1)], $$
$$A(0)\approx   4[1-(k_oa)^2(1-(k/k_o)^2/2) \ln (k_oa/\pi ))/(2\pi)]
/[(k_oa)^2/(k^2/k_o^2-1)].$$ 

Similarly for $n\neq 0,$
in the limit $k_0R<<1$, $J_n(x)\approx (x/2)^n/n![1-(x/2)^2/(n+1)]$, 
$J'(x)\approx (n/x)(x/2)^n/n![1-(n+2)(x/2)^2/(n(n+1))]$,
$N_n(x)=-(2/x)^n(n-1)![1+n(x/2)^2]/\pi.$ 
$$
S(n)\approx -2\pi [i\exp(i\phi_k)]^n  
x[d/dx(J_n(xk/k_o))H_n(x)-J_n(xk/k_o)d/dx(H_n(x))]/[(k_oa)^2(k^2/k_o^2-1)]|_{x_i} 
$$
The dominant contribution in the long wavelength limit is obtained
by replacing H by iN.
$$
A(n)\approx 4 [i\exp(i\phi_k)k/k_o]^n [1+O(k_o^2a^2)]/[(k_oa)^2(k^2/k_o^2-1)] 
$$


\end{references}
\end{document}